\documentclass[a4paper,10pt,twoside]{cpc-hepnp}

\RequirePackage{graphicx}
\RequirePackage{amsmath}
\begin{document}

\title{Measurements of the effect from a hidden photon dark matter using a multi-cathode counter technique}

\author{%
\quad A.V.Kopylov $^{1}$\email{beril@inr.ru}%
\quad I.V.Orekhov \quad V.V.Petukhov }

\maketitle

\address{%
Institute for Nuclear Research of the Russian Academy of Sciences, \\
Moscow, 60-th October Anniversary prospect, 7A,  117312,  Russia}

\begin{abstract}
The results of the experiment to search for hidden photons using a
multi-cathode counter technique are presented. The new results of
measurements using a counter filled with $Ne + CH_4$ (10\%) are
given. The analysis of the data obtained is provided and future
perspectives are outlined.
\end{abstract}



\section{Introduction}
\label{intro} To unveil the mystery of dark matter, researchers make
great efforts in detecting the effects from the constituents of dark
matter. Many different techniques are used in these efforts. We have
developed a new method: a multi-cathode counter to measure the rate
of emission of single electrons from a cathode of a proportional
counter that can be attributed to the effect from hidden photons
(HPs) of cold dark matter (CDM)~\cite{Ref1}, ~\cite{Ref2}. A very
substantial feature of this method is that free electrons of a
degenerate electron gas of a metal are used as a target for HPs, and
the expected effect is proportional to the surface of a cathode of a
counter. This makes our technique different from the volume
detectors that use valence electrons~\cite{Ref3} as a target. This
could be very substantial, accounting for a fact that there is still
no strong theory on HPs. Our detector registers single electrons
emitted from a cathode of a counter, and it has been calibrated by
UV light to ensure that it is very sensitive to the energy of a HP
of about a few eV, provided this energy is above the work function
of a metal of a cathode of a counter. Our detector was tested in
exactly this energy range that is very important for the reliability
of the obtained data. In interpreting the results of any experiment
the important question should be addressed whether the detector has
been calibrated in the energy range for which the interpretation has
been made. Otherwise the interpretation may be not valid.

\section{Methods}
\label{methods} A schematic of the multi-cathode counter is shown in
figure 1.

\begin{figure}[htb]
\begin{center}
\includegraphics[width=8cm]{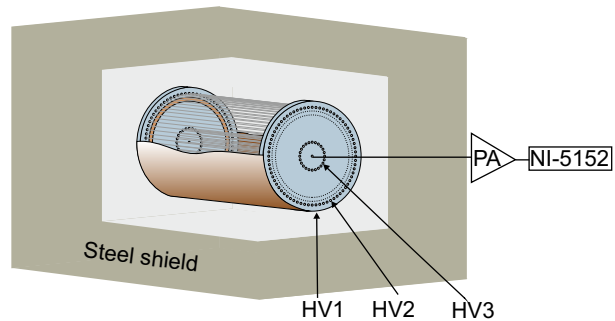}
\vspace{0.6cm} \caption{\label{fig1} Schematic of the multi-cathode
counter. HV1, HV2, HV3 - potentials at cathodes 1, 2 and 3, PA -
preamplifier, NI-5152 - 8 bit digitizer}
\end{center}
\end{figure}

The effect measured by this technique is determined by the
difference $R_{MCC}$ of two count rates~\cite{Ref1}: the one
measured in the Configuration 1 when electrons emitted from a
cathode of the counter reach a central wire (anode) of the counter
and the other in the Configuration 2 when the electrons emitted from
a cathode are scattered back by the locking negative potential of
the second cathode near the first one. If we assume that the
emission of the electron is the result of the conversion of a dark
photon with mass (energy) $m_{\gamma'}$, then the power that can be
prescribed to this rate $R_{MCC}$ can be expressed as

\begin{equation}
\label{eq:1} P=2\alpha^{2}\chi^{2}\rho_{CDM}A_{cath}
\end{equation}

\noindent where $\alpha=cos(\theta)$ and $\theta$ is the angle
between the direction of the HP field and surface of the cathode. If
the HP vector is distributed randomly, then $\alpha^2 = 2/3$;
$\rho_{CDM} \approx 0.3 GeV/cm^3$ is the energy density of CDM that
we assume to be equal to the energy density of the HPs; $A_{cath}$
is the area of the cathode of the counter, and $\chi$ is a
dimensionless parameter that quantifies the kinetic mixing, as
explained in~\cite{Ref4}. If dark matter is composed entirely of
HPs, then this power is

\begin{equation}
\label{eq:2} P=m_{\gamma'} \cdot R_{MCC} /\eta
\end{equation}

\noindent where $\eta$ is the quantum efficiency for a photon with
energy $m_{\gamma'}$ to yield a single electron from the surface of
the metal. By combining equations (1) and (2) we obtain

\begin{equation}
\label{eq:3}
\begin{split}
\chi_{sens}=2.9\cdot10^{-12}\left(\frac{R_{MCC}}{\eta\cdot 1
Hz}\right)^{\frac{1}{2}}\left(\frac{m_{\gamma'}}{1
eV}\right)^{\frac{1}{2}}\left(\frac{0.3\,{\rm
GeV/cm^3}}{\rho_{CDM}}\right)^{\frac{1}{2}}\left(\frac{1
m^{2}}{A_{MCC}}\right)^{\frac{1}{2}}\left(\frac{\sqrt{2/3}}{\alpha}\right)
\end{split}
\end{equation}

It is assumed herein that the quantum efficiency $\eta$ for the
conversion of the HPs at the surface of the metal cathode is equal
to that of a real photon of the same energy.

The apparatus and data analysis have been described in~\cite{Ref1},
~\cite{Ref2}. The results obtained with a counter with copper and
aluminum cathodes filled with an $Ar + CH_4$ (10\%) mixture at 1 bar
have been presented in~\cite{Ref2}. Herein, we present new results
of our measurements during the 111 days obtained with a counter with
an aluminum cathode filled with a mixture of $Ne + CH_4$ (10\%) at 1
bar. Using this mixture, it was possible to decrease the high
voltage applied to the Cathodes 1 and 2 by approximately 450 V,
which resulted in the decrease of leakage current. Thus, we produced
a substantial improvement in the work of the counter and obtained
the lowest result for $r_{MCC} = R_{MCC}/A_{cath} = (- 0.33 \pm
0.7)\cdot 10^{-6} Hz/cm^2$.

\section{Results and discussion}
\label{results} The results of measurements in the Configurations 1
and 2 are presented at figure 2.

\begin{figure}[htb]
\begin{center}
\includegraphics[width=8cm]{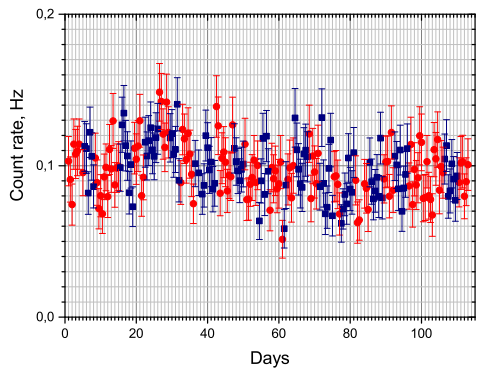}
\figcaption{\label{fig2} The count rates measured in the
Configurations 1 (circles) and 2 (squares). The temperature of the
counter varied in the region from $5^{o}$C to $15^{o}$C}
\end{center}
\end{figure}

The limits obtained from the analysis of these data are presented in
the figure 3. As can be seen from this figure, we have achieved some
progress with a mixture of $Ne + CH_4$ (10\%). This result is still
within the limits determined by the luminosity of the Sun, but we
should note that the physics of the processes in the interior of the
stars with high temperature and density plasma, still unattainable
in terrestrial laboratories, may be very different from the physics
in our case.

\begin{figure}[htb]
\begin{center}
\includegraphics[width=8cm]{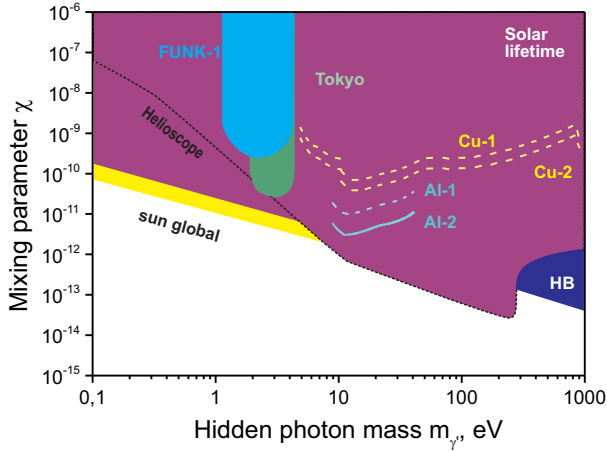}
\figcaption{\label{fig3} Limits at a 95\% CL obtained from the
series of measurements Cu-1, Cu-2 and Al-1~\cite{Ref2} and also Al-2
with a mixture $Ne + CH_4$ (10\%). Here the Tokyo limits are
from~\cite{Ref5} and the FUNK-1 limits from~\cite{Ref6}}
\end{center}
\end{figure}

Our limits are also higher than the ones obtained with the volume
detectors~\cite{Ref3}, but as emphasized earlier, we used the free
electrons of a degenerate electron gas as a target in our detector,
whereas, the target in volume detectors are valence electrons. The
physics of these two cases may be very different. Further
improvements can be achieved cooling the detector until lower
temperatures diminish the contribution of thermionic noise from the
impurities on the surface of the wires. We are also developing a new
detector with a nickel cathode, as it has a relatively high work
function.

\vspace{3mm}

{\bf Acknowledgments} The work was supported by the basic research
program of INR RAS. The authors would like to thank Enago
(www.enago.com) for the English language review.

\clearpage
\end{document}